\documentclass[aps,prd,twocolumn,nofootinbib,superscriptaddress]{revtex4}

\usepackage{amsmath}
\usepackage{amssymb}
\usepackage{graphicx}
\usepackage{color}
\usepackage{multirow}

\newcommand{\be}{\begin{equation}}
\newcommand{\ee}{\end{equation}}

\newcommand{\df}{\text{d}}

\newcommand{\wt}[1]{\widetilde{#1}}
\newcommand{\ord}[1]{\mathcal{O}\!\left(#1\right)}

\DeclareRobustCommand{\Sec}[1]{Sec.~\ref{#1}}

\DeclareRobustCommand{\Fig}[1]{Fig.~\ref{#1}}

\DeclareRobustCommand{\Eq}[1]{Eq.~(\ref{#1})}
\DeclareRobustCommand{\Eqs}[2]{Eqs.~(\ref{#1}) and (\ref{#2})}
\DeclareRobustCommand{\Ref}[1]{Ref.~\cite{#1}}

\newcommand{\nn}{\nonumber}

\begin{document}

\title{Integrated and Differential Accuracy in Resummed Cross Sections}

\author{Daniele Bertolini}
\affiliation{Ernest Orlando Lawrence Berkeley National Laboratory, University of California,
Berkeley, CA 94720, U.S.A.}
\affiliation{Berkeley Center for Theoretical Physics, University of California,
Berkeley, CA 94720, U.S.A.}

\author{Mikhail P. Solon}
\affiliation{Ernest Orlando Lawrence Berkeley National Laboratory, University of California,
Berkeley, CA 94720, U.S.A.}
\affiliation{Berkeley Center for Theoretical Physics, University of California,
Berkeley, CA 94720, U.S.A.}

\author{Jonathan R. Walsh}
\affiliation{Ernest Orlando Lawrence Berkeley National Laboratory, University of California,
Berkeley, CA 94720, U.S.A.}
\affiliation{Berkeley Center for Theoretical Physics, University of California,
Berkeley, CA 94720, U.S.A.}

\begin{abstract}
Standard QCD resummation techniques provide precise predictions for the spectrum and the cumulant of a given observable. The integrated spectrum and the cumulant differ by higher-order terms which, however, can be numerically significant. In this paper we propose a method, which we call the $\sigma\text{-improved}$ scheme, to resolve this issue. It consists of two steps: (i) include higher-order terms in the spectrum to improve the agreement with the cumulant central value, and (ii) employ profile scales that encode correlations between different points to give robust uncertainty estimates for the integrated spectrum. We provide a generic algorithm for determining such profile scales, and show the application to the thrust distribution in $e^+e^-$ collisions at NLL$'$+NLO and NNLL$'$+NNLO.
\end{abstract}

\maketitle

\section{Introduction}

Quantum chromodynamics (QCD) is essential for understanding data from collider experiments. Countless measurements at the Large Hadron Collider (LHC), from Higgs coupling measurements to new physics searches, rely on precision QCD predictions. The success of these programs has been enabled through remarkable advances in the community's ability to calculate cross sections with a level of precision that keeps pace with constantly improving experimental measurements. This will continue to be the case for the remainder of the LHC program and for future colliders.

This paper focuses on predictions for observables in QCD that require resummation of large logarithms. Such observables are standard at collider experiments, from event and jet shape observables to classical observables like $q_T$, the transverse momentum of the vector boson in Drell-Yan production. The most precise calculations of these observables match resummed and fixed order results to obtain an accurate prediction across the entire range of the observable. However, there is a common inconsistency in resummed predictions, one which we address in this work.

Resummed calculations for a generic observable $\tau$ make two predictions: the spectrum $\df\sigma/ \df\tau$ (cross section differential in $\tau$), and the cumulant $\Sigma (\tau)$ (cross section integrated over $\tau$). Using standard resummation techniques, these predictions differ by higher-order terms (see, e.g., \Ref{Almeida:2014uva}),
\be \label{inconsistency}
\int_0^{\tau} \!\df\tau' \, \frac{\df\sigma}{\df\tau'} = \Sigma(\tau) + \text{higher order}\,,
\ee
which can be numerically significant. An equivalent form of the inconsistency is that the derivative of the cumulant is inconsistent with the spectrum: $\df\sigma/\df\tau = \df\Sigma(\tau) / \df\tau + \text{higher order}$.  

Each prediction is internally consistent and valid: the spectrum accurately predicts the value of the differential cross section and its uncertainties point-by-point in $\tau$, while the cumulant accurately predicts the integrated cross section and its uncertainties.  However, standard resummation methods do not accurately model the long-range correlations in the spectrum, and when the spectrum is integrated this leads to the inconsistency with the cumulant.  Taking $\tau\to\infty$ in \Eq{inconsistency}, a simple statement is that the integral of the spectrum does not give the correct inclusive cross section and its uncertainties at the relevant fixed order accuracy. On the other hand, while these quantities are correctly predicted by the cumulant, it is a poor model of the short-range uncertainty correlations, and so its derivative fails to accurately predict the point-by-point uncertainties in the spectrum. We will resolve this basic problem, making the spectrum and cumulant predictions consistent.

The inconsistency in \Eq{inconsistency} arises from the fact that the renormalization and factorization scales are chosen (by necessity) to be $\tau$-dependent.  We will show that a simple constraint on these $\tau$-dependent scales will render the spectrum and cumulant consistent, allowing the spectrum to correctly predict the inclusive cross section and its uncertainties. We will provide a generic algorithm, which we call Bolzano's algorithm, to choose scales that satisfy this constraint.

The layout of this paper is as follows.  In \Sec{sec:formalism}, we discuss the spectrum and cumulant predictions and the source of the inconsistency in \Eq{inconsistency} in detail.  In \Sec{sec:Bolzano}, we present a technique to make the spectrum and cumulant consistent, and in \Sec{sec:thrust} we implement the solution for the example of the thrust distribution in $e^+e^-$ collisions. We conclude in \Sec{sec:conclusions} and describe a specific implementation of the algorithm in Appendix~\ref{app:Bolzano}.

\section{The Resummed Spectrum, Cumulant, and their Uncertainties}
\label{sec:formalism}

\begin{figure*}[t!]
\begin{center}
\includegraphics[width=0.48\textwidth]{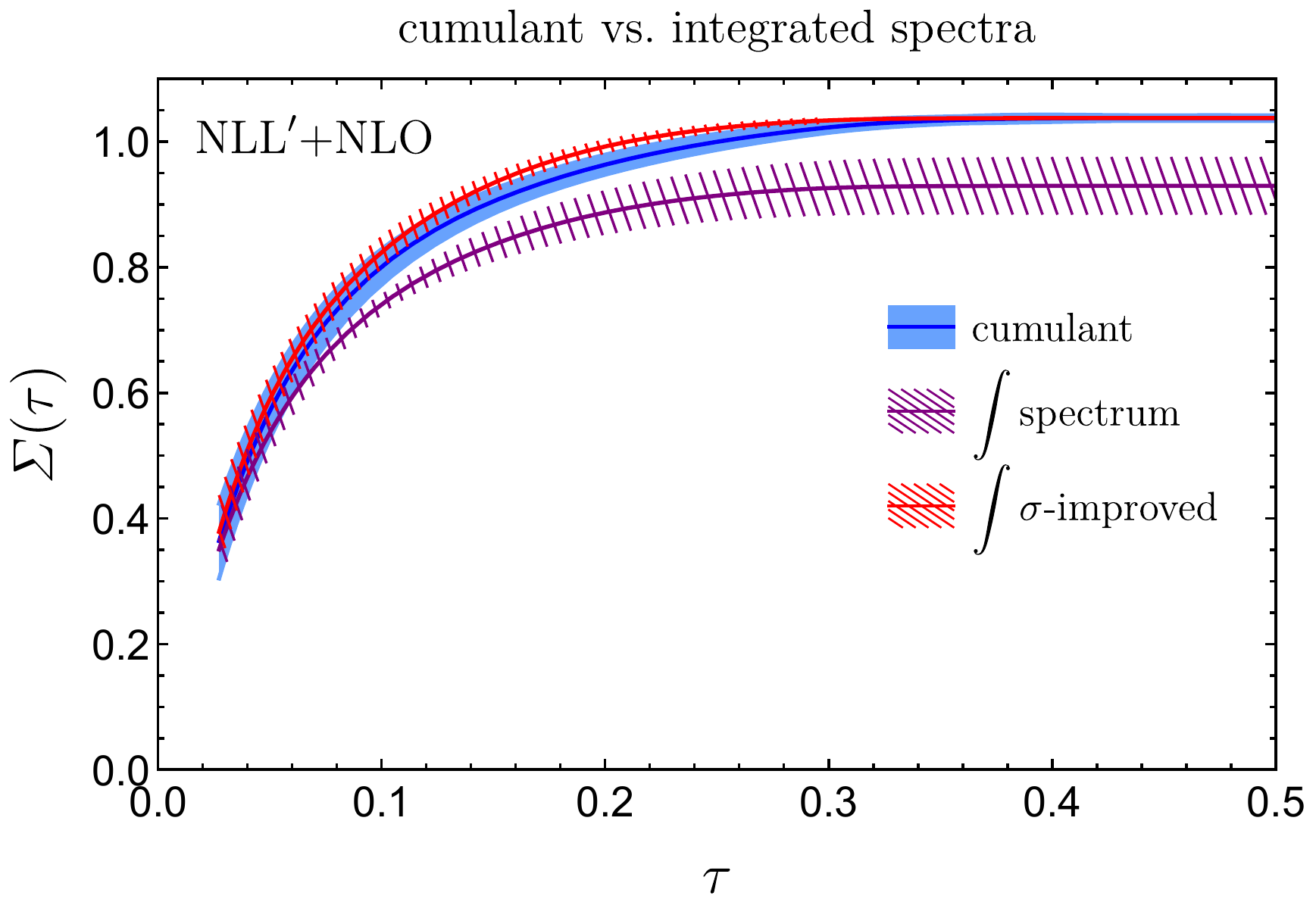}
\hfill
\includegraphics[width=0.48\textwidth]{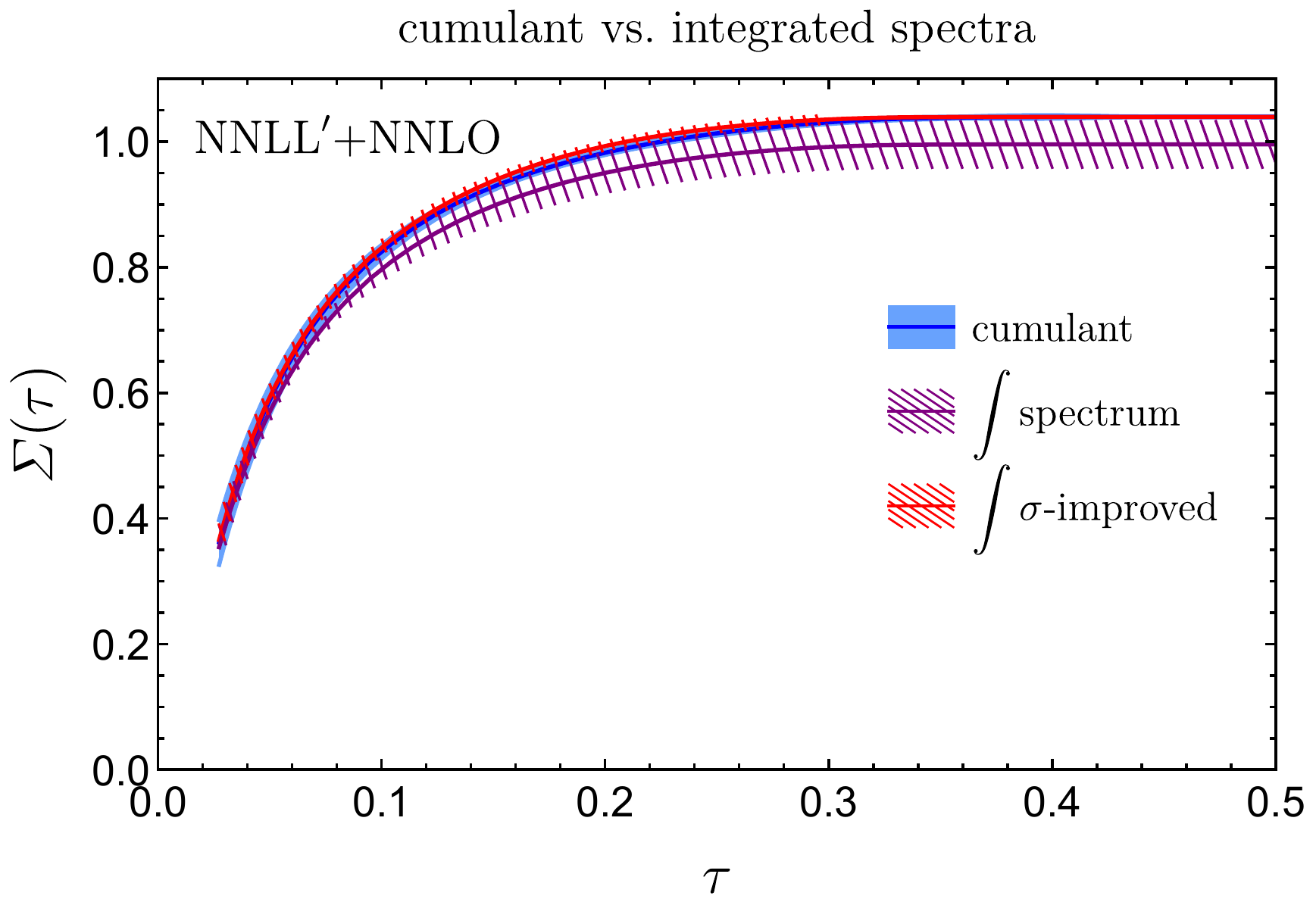}
\caption{Comparison of the cumulant with the integrated resummed spectrum for thrust in $e^+e^-$ collisions at NLL$'$+NLO and at NNLL$'$+NNLO. The cumulant matches onto the inclusive cross section at large $\tau$, while the standard integrated spectrum differs in both value and uncertainty. The integrated spectrum with the $\sigma\text{-improved}$ scheme gives a consistent prediction.}
\label{fig:cumulantcompareSTD}
\end{center}
\end{figure*}

Resummed calculations generally operate within one of two frameworks: soft-collinear effective theory (SCET)~\cite{Bauer:2000ew, Bauer:2000yr, Bauer:2001ct, Bauer:2001yt}, or direct QCD (dQCD) (see, e.g., Refs.~\cite{Catani:1992ua,Contopanagos:1996nh}).  Both frameworks provide equivalent predictions, built by factorizing the cross section in terms of more universal matrix elements that depend on a restricted set of dynamics and scales~\cite{Almeida:2014uva,Sterman:2013nya,Bonvini:2013td,Bonvini:2012az}.  The scales at which the factorization occurs are handles by which uncertainties can be assigned; this process is called scale variation.  There are many schemes to perform scale variation, and the assessment of uncertainties can be subjective.  However, by comparing different orders of the calculation (e.g. NLL$'$+NLO versus NNLL$'$+NNLO) to get a sense of convergence, one can judge the robustness of an uncertainty scheme.

One feature common to factorization theorems for resummed calculations is that some factorization scales are naturally observable-dependent.  For example, if we have a resummed cross section for an $e^+e^-$ dijet event shape $\tau$, the factorization theorem for the spectrum in SCET has the form
\begin{align} \label{factthm}
\frac{\df\sigma}{\df\tau} &= H (Q, \mu_H) U_H (\mu_H, \mu_0) \bigl[ J(Q, \tau, \mu_J) U_J (\mu_J, \mu_0) \bigr] \nn\\
& \quad \otimes \bigl[ J(Q, \tau, \mu_J) U_J (\mu_J, \mu_0) \bigr] \otimes \bigl[ S(\tau, \mu_S) U_S (\mu_S, \mu_0) \bigr] \,,
\end{align}
where $Q$ is the center of mass energy, and $H$, $J$, and $S$ are the hard, jet, and soft functions with corresponding evolution factors $U_H$, $U_J$, and $U_S$, and factorization scales $\mu_H$, $\mu_J$, and $\mu_S$~\cite{Bauer:2008dt,Hornig:2009vb,Lee:2006nr}.  The evolution factors sum the large logarithms of the factorization scales to the arbitrary common scale $\mu_0$, and the factorization scales are chosen to be similar to the ``natural'' scales in the functions.  In this example, this means that the jet and soft factorization scales will be $\tau$-dependent; we refer to them as profile scales (see, e.g., Refs~\cite{Ligeti:2008ac,Abbate:2010xh,Berger:2010xi,Kang:2013nha,Gritschacher:2013pha,Chang:2013iba,Jain:2012uq,Alioli:2012fc,Bauer:2011uc,Jouttenus:2013hs,Liu:2013hba} for a discussion of profile scales in various contexts).  A construction in dQCD will give the same essential features (see, e.g., Refs.~\cite{Berger:2003iw,Lee:2006nr}).

The spectrum and cumulant predictions for a given observable each have de facto features guaranteed by their definitions. For the spectrum, robust matching to fixed order calculations in the large $\tau$ limit (where logarithms are no longer large) will yield accurate predictions for the distribution.  This feature is nontrivial, as the singular (which is resummed) and nonsingular (which is typically not) components of the spectrum have large cancellations at large $\tau$, and the matching must preserve these cancellations.  This is tantamount to the statement that one must be careful when, and how, the resummation is turned off at large $\tau$.  Additionally, robust uncertainty models will yield accurate uncertainty estimates point-by-point in $\tau$.  

For the cumulant, consistent matching to fixed order predictions will ensure the following condition is met:
\begin{align}
\Sigma (\tau\to \infty ; \mu_i) \to \sigma_{\rm incl} (\mu) \,,
\end{align}
where $\mu_i$ represents the set of factorization scales and $\mu$ is the renormalization scale (or represents the renormalization and factorization scales for hadronic collisions).  That is, the fixed order inclusive cross section is recovered in the large $\tau$ (inclusive) limit, which is precisely where fixed order predictions are robust.  
This implies that the cumulant is accurately modeling the long-range scale uncertainties in the cross section, while the spectrum is accurately modeling the short-range scale uncertainties.

These features of the spectrum and cumulant become inconsistent because of the $\tau$-dependence in the factorization scales (see, e.g., \Ref{Almeida:2014uva}). Consider the spectrum and cumulant before scales are chosen: by construction they obey the relation
\begin{align}
\frac{\partial}{\partial \tau} \Sigma(\tau; \mu_i) = \frac{\df\sigma}{\df\tau} (\mu_i) \,.
\end{align}
Because the only $\tau$ dependence is through the explicit $\tau$, we can convert the partial derivative to a full derivative without penalty.  However, once we choose factorization scales to be $\tau$-dependent (e.g., $\mu_J$ and $\mu_S$), we have
\begin{align}
\frac{\df}{\df\tau} = \frac{\partial}{\partial \tau} + \frac{\df\mu_J}{\df\tau} \frac{\partial}{\partial \mu_J} + \frac{\df\mu_S}{\df\tau} \frac{\partial}{\partial \mu_S} \,.
\end{align}
This introduces a difference between the cumulant derivative and the spectrum:
\begin{align} \label{cumulantspectrumdiff}
&\frac{\df}{\df\tau} \Sigma(\tau; \mu_i) - \frac{\df\sigma}{\df\tau} (\mu_i) 
= \sum_{F = J,S} \frac{\df\mu_F}{\df\tau} \frac{\partial}{\partial \mu_F} \Sigma(\tau; \mu_i) \,.
\end{align}
Because the scale dependence cancels through the resummation order achieved, this difference is strictly higher order.  This nonzero difference shows explicitly that the integral of the spectrum is not equal to the inclusive cross section; rather it is
\begin{align}
\int \df\tau \frac{\df\sigma}{\df\tau} (\mu_i) = \sigma_{\rm incl} (\mu) - \sum_{F = J,S} \int \df\mu_F \, \frac{\partial}{\partial \mu_F} \Sigma(\tau; \mu_i) \,.
\end{align}
This last term can be numerically significant even though it is higher order, as it accumulates over the entire spectrum.  Furthermore, the value of this term will generally vary for different scale variations, implying that the uncertainty of the integrated spectrum can also be different from the fixed order value.

In \Fig{fig:cumulantcompareSTD}, we compare the cumulant to the integrated spectrum for thrust in $e^+e^-$ collisions, at both NLL$'$+NLO and NNLL$'$+NNLO~\cite{Farhi:1977sg,Fleming:2007qr,Schwartz:2007ib}. The plots illustrate the discrepancy between the two predictions: the integrated spectrum does not match the cumulant in central value or uncertainty over most of the range in $\tau$. In particular, at large $\tau$, the integrated spectrum does not match the inclusive cross section or its uncertainties.

The uncertainties are estimated through two types of scale variations: those that probe the size of the logarithms being resummed (resummation variations), and those that probe the absolute size of the scales, including the renormalization scale (fixed order variations). Resummation variations probe the size of the logarithms of scale ratios by varying the profile scales. For the case of thrust, e.g., the logarithms of scale ratios are of the form
\be
\ln \frac{\mu_H^2}{\mu_J^2} \,, \qquad \ln \frac{\mu_J^2}{\mu_S^2} \,,
\ee
and there is a canonical relationship between the scales, $\mu_J^2 = \mu_H \, \mu_S$, that can be used to define the $\mu_J$ profile in terms of the $\mu_S$ profile.  One may choose the following resummation scale variations:
\begin{itemize}
\item Vary $\mu_S$ by a factor $f_S (\tau)$ (and its inverse).
\item Vary $\mu_J$ by a factor $f_J (\tau)$ and $\mu_S$ by $f_J^2 (\tau)$, keeping the canonical relationship intact.
\end{itemize}
Additionally, one may choose to vary the profile shapes to quantify the uncertainty associated with the choice of profile scales.  For the fixed order variations, a standard procedure is to vary all scales by a common factor of $2$ or $1/2$.  This maintains the size of the logarithms but varies the renormalization scale dependence in the resummation as well as the matching.

\section{A Spectrum with Integrated Accuracy}
\label{sec:Bolzano}

In this section we describe the general approach used in this work to obtain consistent predictions for the spectrum and cumulant. As discussed above, these quantities have complementary advantages and disadvantages, and our method is a way of capturing the advantages of both. It consists of two steps:
\begin{itemize}
\item[1.] Add higher-order terms to the resummed spectrum to improve the agreement with the inclusive cross section.
\item[2.] Asses the resummation uncertainty using profile scales that preserve the integrated value of the spectrum.
\end{itemize}

The first step resolves the numerical difference between the integrated spectrum and the cumulant, which can be outside the uncertainties of either prediction (see \Fig{fig:cumulantcompareSTD}). The higher-order terms are also useful to ensure the spectrum matches the inclusive cross section for the central, up, and down scale variations (each with a different renormalization scale $\mu$). 

The second step allows for resummation variations to be fully uncorrelated with the uncertainty in the inclusive cross section, which is governed by fixed order variations. Note that this is not guaranteed using standard profile scales (as one can see in \Fig{fig:cumulantcompareSTD}, standard profile  variations lead to large uncertainties in the integrated spectrum). We have devised an algorithm to generate such cross section-preserving profile scales, requiring them to satisfy basic criteria such as monotonicity, smoothness, and boundedness.

The following subsections describe these two steps in further detail.

\subsection{Step 1: Higher-Order Terms}
\label{subsec:higherorder}

We add the following higher-order terms to the standard resummed spectrum:
\begin{align} \label{eq:higherorder}
\delta\sigma_R (\tau; \wt{\mu}_i) = \kappa(\tau) \Bigl(\frac{\df}{\df\tau} \Sigma(\tau; \wt{\mu}_i) - \frac{\df\sigma}{\df\tau} (\wt{\mu}_i) \Bigr) \,,
\end{align}
which restore the inclusive cross section in the spectrum,
\be
\int \df \tau \Bigl[ \frac{\df\sigma}{\df\tau} (\mu_i) + \delta\sigma_R (\tau; \wt{\mu}_i) \Bigr] = \sigma_{\rm incl} (\mu) \,.
\ee
Above, the $\wt{\mu}_i$ are special profiles with two features.  First, since the difference between the spectrum and the derivative of the cumulant is proportional to $\df\wt{\mu}_i / \df\tau$, we want these profiles to have smooth derivatives.  Second, they are designed to turn off the resummation earlier than the standard profiles.  This ensures that the cancellation between the nonsingular and resummed singular parts of the matched spectrum are preserved in the tail region of the distribution.  

The function $\kappa(\tau)$ is a smooth function of $\tau$ that goes to zero at large $\tau$ and whose maximum is an $\ord{1}$ value. Since $\kappa(\tau)$ enables us to tune the effect of the higher order terms, we will take them to be such that the integral of the spectrum exactly matches the inclusive cross section for all fixed order scale variations (variations of the matching scale).  This ensures that the fixed order variations preserve the inclusive cross section and its uncertainties.
Of course, tuning the higher order terms to precisely match the inclusive cross section is not necessary, but it simplifies the latter step of producing profiles that preserve the inclusive cross section (e.g., it allows the straightforward identification of the central scale).

Including the higher-order terms, the spectrum is
\begin{align} \label{eq:sigmaimproved}
\frac{\df\sigma_R}{\df\tau} (\mu_i) = \frac{\df\sigma}{\df\tau} (\mu_i) + \delta\sigma_R (\tau; \wt{\mu}_i) \,,
\end{align}
where $\df\sigma/\df\tau$ is the standard resummed spectrum (without any higher-order terms added). We will call $\df\sigma_R/\df\tau$ the $\sigma\text{-improved}$ spectrum. 
A similar procedure was adopted in \Ref{Alioli:2015toa}.
For the thrust example discussed in \Sec{sec:thrust}, we give the explicit form of $\widetilde{\mu}_i$ and $\kappa(\tau)$ in Appendix~\ref{app:Bolzano}.

\subsection{Step 2: Bolzano's Algorithm}
\label{subsec:Bolzano}

Finding profile scales $\mu_i$ that give a spectrum whose integral is the inclusive cross section can be phrased in terms of solving an integral equation:
\be \label{eq:main}
\int\df\tau \frac{\df\sigma_R}{\df\tau} (\mu_i) = \sigma_{\rm incl}\,,
\ee
where the renormalization scale dependence of the inclusive cross section is implicit.
Given the complex dependence of the spectrum on the factorization scales, an analytic approach is not feasible but we can devise a numeric algorithm to find profile scales that solve the equation to within a negligible tolerance.  We will discuss the algorithm in terms of finding profile scales $\mu_S (\tau)$, but the same ideas carry through straightforwardly for $\mu_J (\tau)$. 

We will identify profile scales obeying the following constraints:
\begin{itemize}
\item $\mu_S(\tau)$ is monotonic and smooth.
\item $\mu_S(\tau)$ has fixed shapes near the endpoints.
\item $\mu_S(\tau)$ is bounded; $\mu_S^{\rm min} (\tau) < \mu_S(\tau) < \mu_S^{\rm max} (\tau)$ .
\end{itemize}

The first condition ensures the smoothness of the spectrum but not necessarily its monotonicity, which should be further checked.

The second condition imposes canonical profile shapes near the endpoints. In the low $\tau$ region, nonperturbative effects on the resummed distribution can be large and often determine the profile scales~\cite{Hoang:2007vb,Korchemsky:2000kp,Korchemsky:1999kt,Hornig:2009vb,Berger:2003pk}. Thus, for $\tau < \tau_\text{NP}$, we fix the profile to a given shape normalized by the value of the profile at $\tau_\text{NP}$. Similarly, in the large $\tau$ region, to preserve the cancellations between singular and nonsingular terms, we fix the shape of profiles for $\tau > \tau_{\rm tail}$.  We have the freedom to change the profile scales in the range $\tau_\text{NP} < \tau < \tau_{\rm tail}$.

The third condition ensures that the profile scales produce reasonable uncertainty estimates, consistent with convergence between different orders of resummed perturbation theory and the relative size of the singular and non-singular contributions. In practice, we choose the boundary functions $\mu_S^{\rm min} (\tau)$ and $\mu_S^{\rm max} (\tau)$ to be the standard minimum and maximum variations from the central profile, so that the goal is to fill the standard band with profiles $\mu_S(\tau)$ that solve \Eq{eq:main}. 

Our strategy to generate profile scales satisfying these constraints is based on the intermediate value theorem (also known as Bolzano's theorem), and we will refer to it as Bolzano's algorithm. Bolzano's theorem states that if a continuous function takes values of opposite sign at the endpoints of an interval, then there is at least one point within the interval where the function vanishes:
\begin{align} \label{eq:Bolzanothm}
\text{continuous} \ f:\,&[a,b]\to\mathbb{R},\,f(a)<0<f(b) \nn \\
&\Rightarrow\exists\,c\in (a,b)\mbox{ such that } f(c)=0 \,.
\end{align}

\begin{figure}[t]
\includegraphics[width=0.48\textwidth]{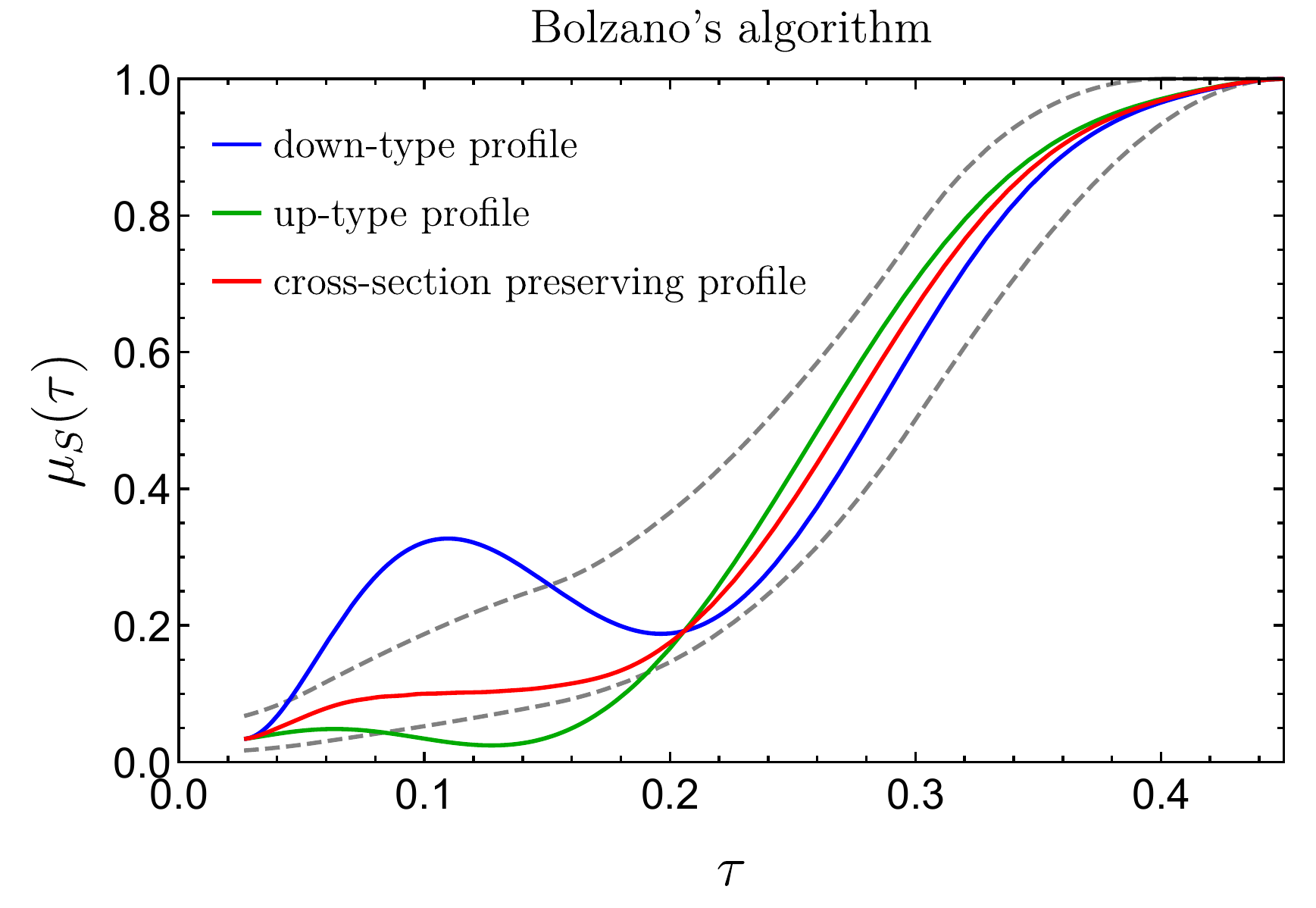}
\caption{Our method to find solutions to \Eq{eq:main} is based on determining profiles that give an inclusive cross section less than (down-type profiles) or greater than (up-type profiles) the true inclusive cross section, and then identifying for each pair the combination that solves the equation and is monotonic, smooth, and bounded.}
\label{fig:BolzanoSketch}
\end{figure}

\begin{figure*}[t!]
\begin{center}
\includegraphics[width=0.48\textwidth]{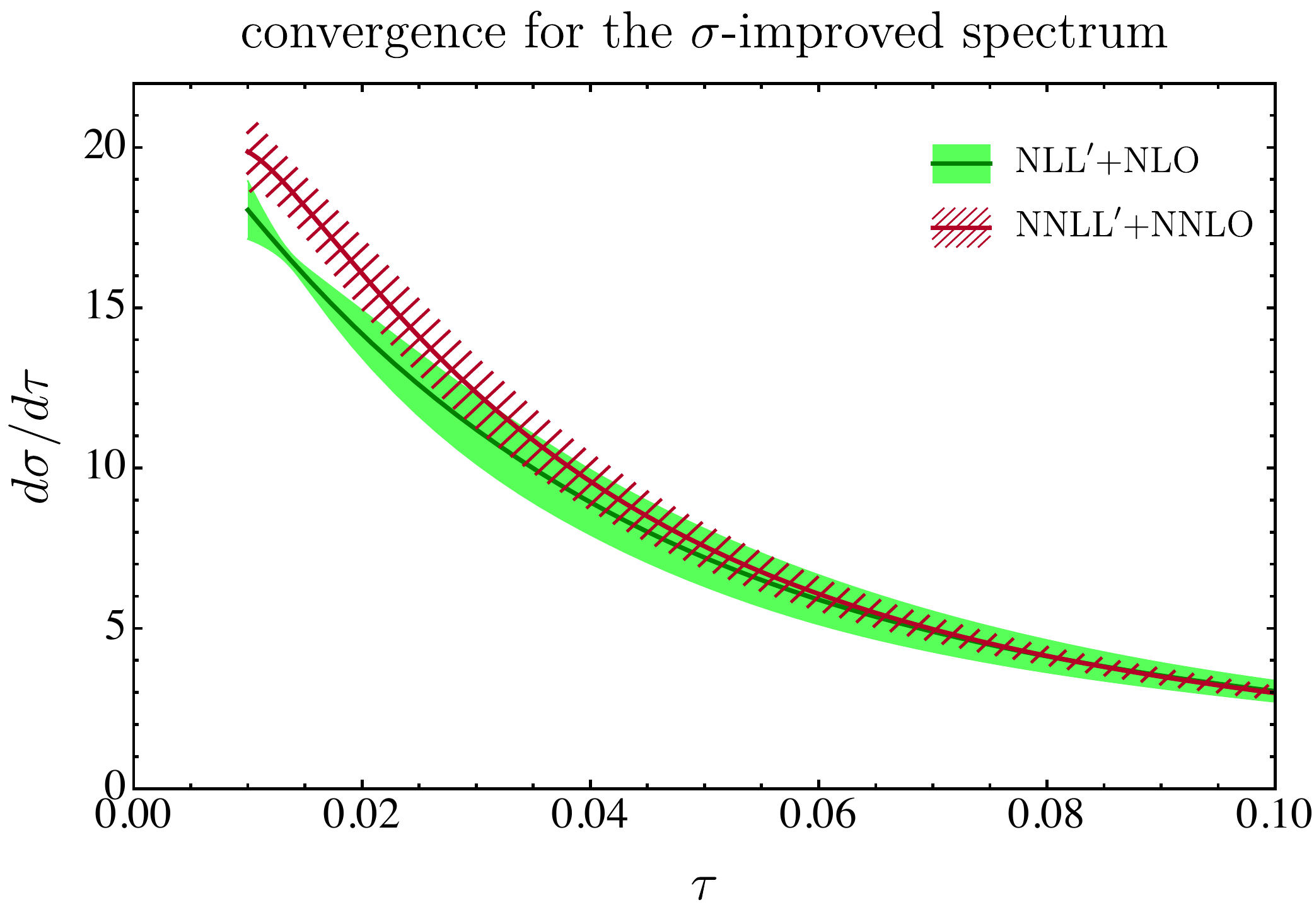}%
\hfill%
\includegraphics[width=0.48\textwidth]{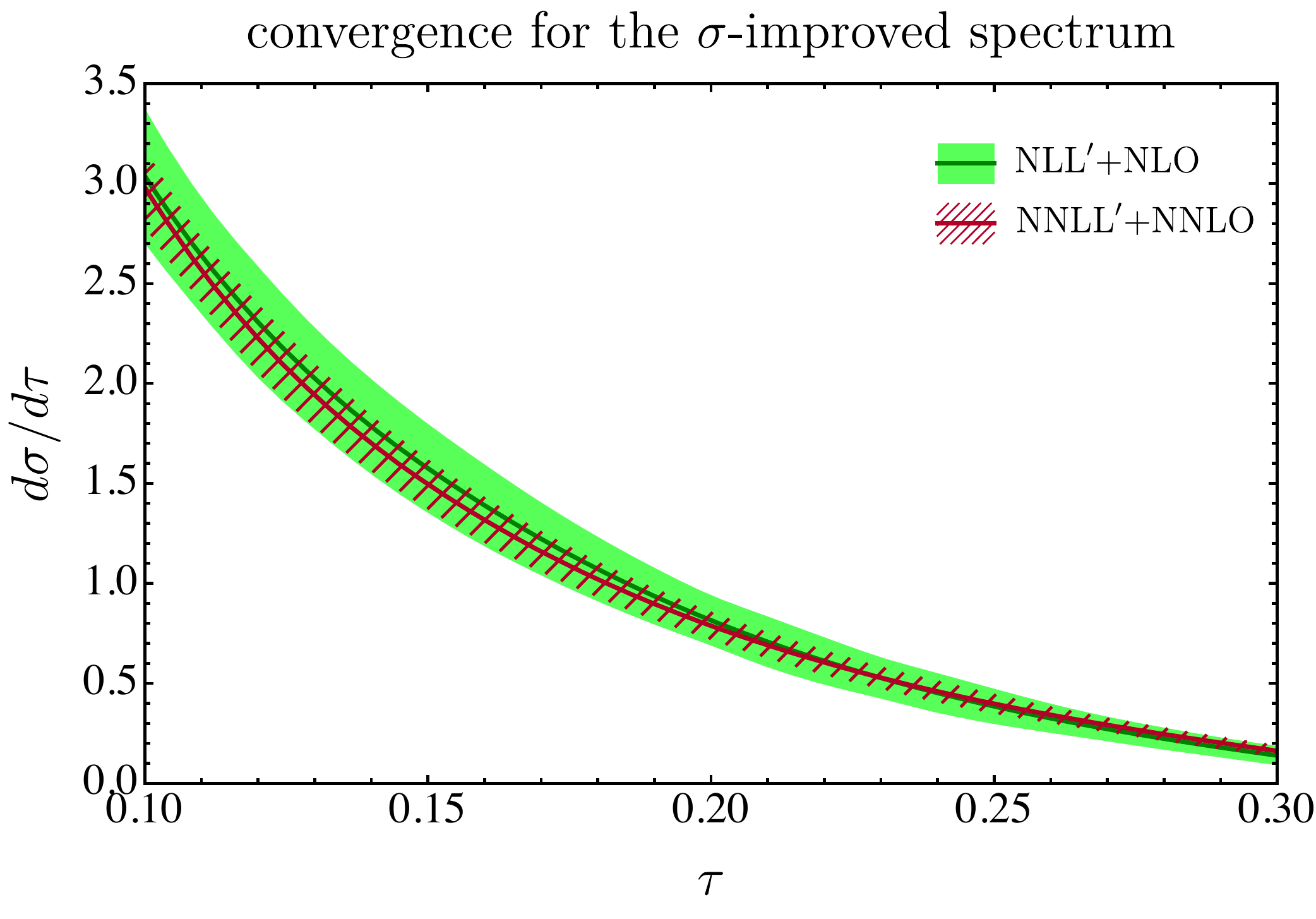}\\
\includegraphics[width=0.48\textwidth]{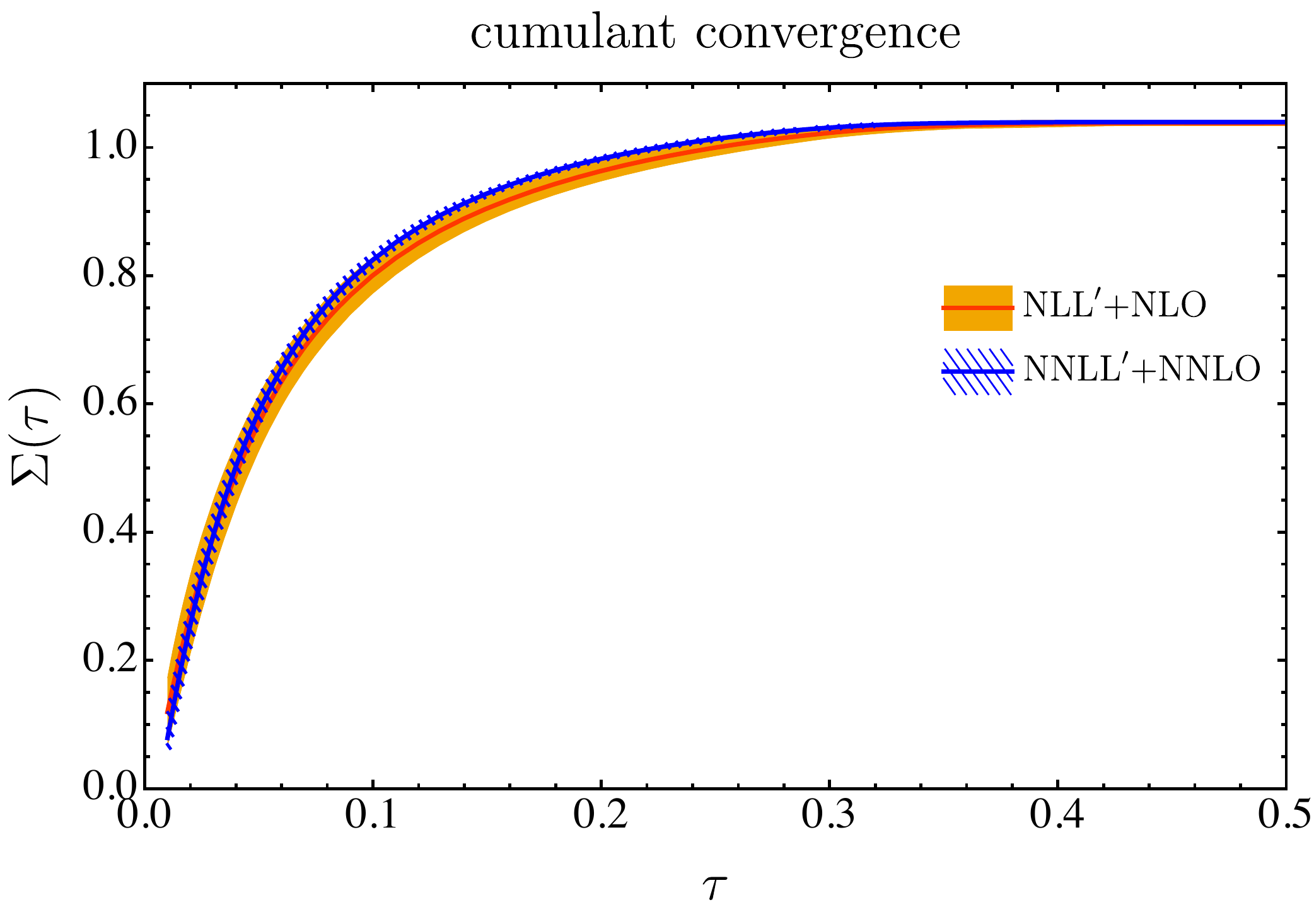}%
\hfill%
\includegraphics[width=0.48\textwidth]{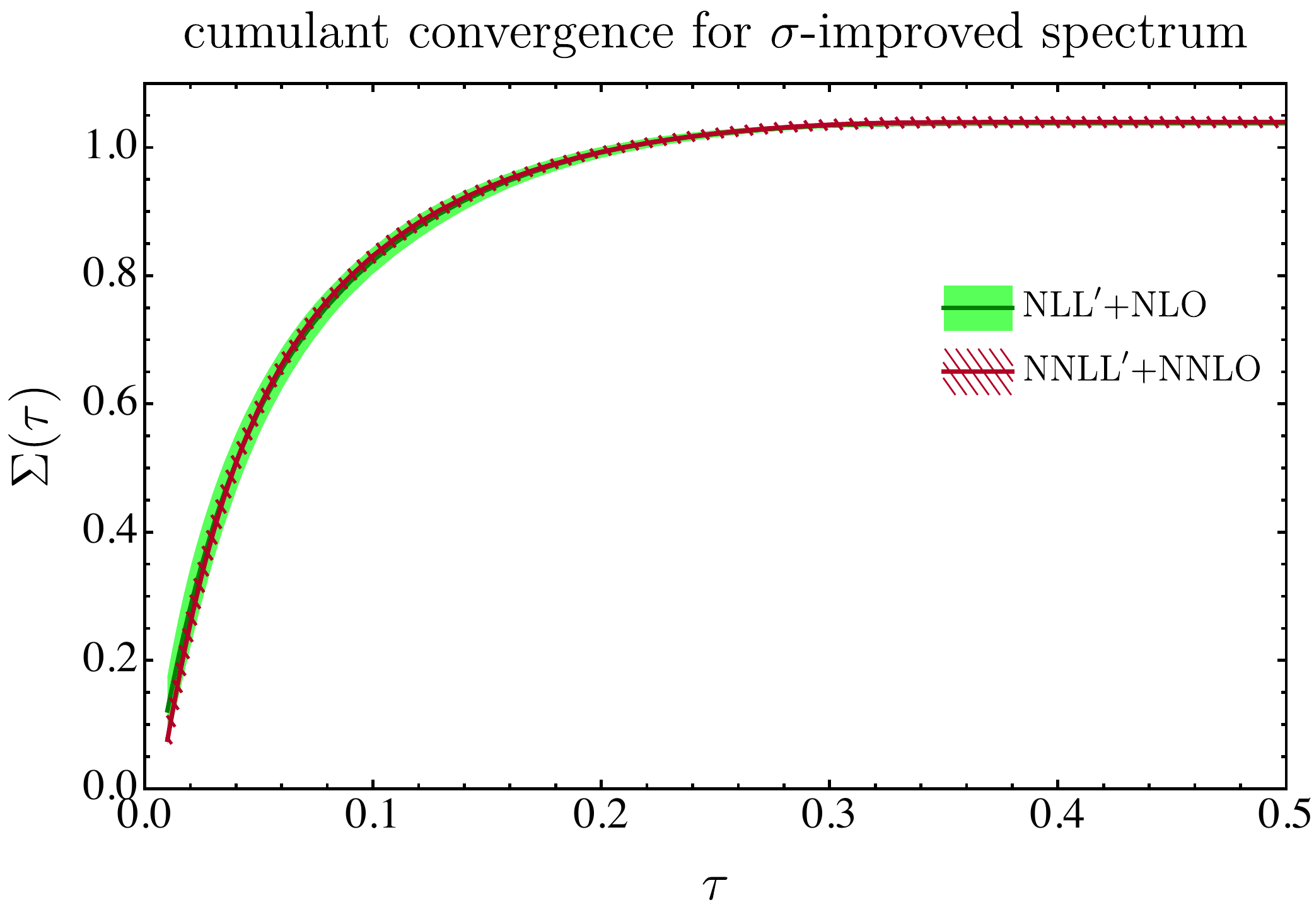}\\
\caption{Convergence of the thrust $\sigma\text{-improved}$ spectrum in the peak (upper left panel) and transition (upper right panel) regions, and of the cumulant (lower left panel) and integrated $\sigma\text{-improved}$ spectrum (lower right panel). The slight non-convergence in the peak region is an artifact of pinching in the resummation scale dependence and exists also in the standard case. In contrast to the standard case, the integrated $\sigma\text{-improved}$  spectrum exhibits the convergence properties of the cumulant, as expected.}
\label{fig:converge}
\end{center}
\end{figure*}

Suppose we have two profiles $\mu_S^d (\tau)$ and $\mu_S^u (\tau)$ which give spectra that integrate to values below and above the inclusive cross section, respectively:
\begin{align} \label{eq:updown}
\int \df \tau \frac{\df\sigma_R}{\df\tau} (\mu_S^d) = \sigma_{\rm incl}^d &< \sigma_{\rm incl}, \nn \\
\int \df \tau \frac{\df\sigma_R}{\df\tau} (\mu_S^u) = \sigma_{\rm incl}^u &> \sigma_{\rm incl} \,.
\end{align}
Then, defining
\begin{align}\label{eq:linearupdown}
\frac{\df\sigma_R^{(d,u)} (\alpha)}{\df\tau} = \alpha \frac{\df\sigma_R}{\df\tau} (\mu_S^u) + (1-\alpha) \frac{\df\sigma_R}{\df\tau} (\mu_S^d) \,,
\end{align}
Bolzano's theorem guarantees that there exists an $\alpha_0 \in [0, 1]$, in this case
\be\label{eq:alpha0}
\alpha_0 = \frac{\sigma_{\rm incl} - \sigma_{\rm incl}^d}{\sigma_{\rm incl}^u - \sigma_{\rm incl}^d} \,,
\ee
such that
\be
\int \df\tau \frac{\df\sigma_R^{(d,u)} (\alpha_0)}{\df\tau} = \sigma_{\rm incl} \,.
\ee
The corresponding profile scale is found by inverting the scale dependence of the spectrum.  

Alternatively, one can also use the same concept to directly solve for profile functions that give a spectrum with the inclusive cross section.  Defining
\be\label{eq:linear2}
\mu_S^{(d, u)} (\alpha) = \alpha \mu_S^u (\tau) + (1-\alpha) \mu_S^d (\tau) \,,
\ee
$\alpha$ smoothly interpolates between the down-type profile $\mu_S^d$ and up-type profile $\mu_S^u$ in \Eq{eq:updown}, and thus there exists an $\alpha_*$ for which
\be
\int \df\tau \frac{\df\sigma_R}{\df\tau} \bigl( \mu_S^{(d,u)}(\alpha_*) \bigr) = \sigma_{\rm incl} \,.
\ee
This approach does not require inverting the spectrum to find the profile, but it requires solving for $\alpha_*$ numerically.

Let us give a formulation of Bolzano's algorithm: 
\begin{itemize}
\item[1.] Generate a set of smooth profiles that have fixed shapes near the endpoints. 
\item[2.] Sort the profiles into down-type and up-type.
\item[3.] For each pair of down-type and up-type profiles, determine the combination whose spectrum integrates to the inclusive cross section.
\item[4.] Select the solutions that are correctly bounded and monotonic.
\item[5.] Define a default central profile, if not assumed to be the standard central profile.
\end{itemize}
This is illustrated in \Fig{fig:BolzanoSketch}. Note that, in the first step, we include profiles that are slightly non-monotonic or slightly outside the bounds since they may still lead to monotonic and bounded solutions. Particular implementations of the algorithm are discussed in Appendix~\ref{app:Bolzano}.
 
\begin{figure}[t]
\includegraphics[width=0.48\textwidth]{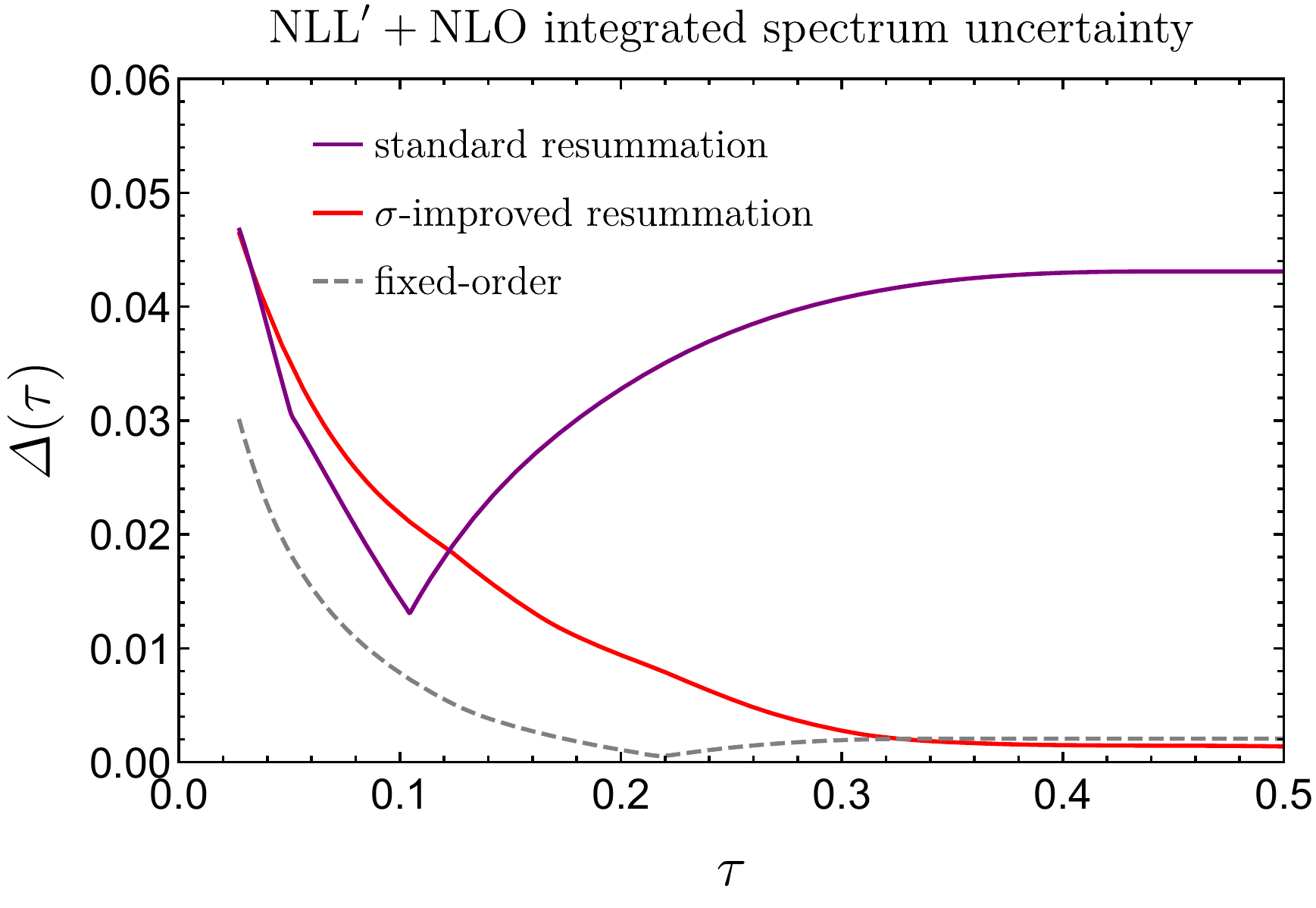}
\caption{Uncertainties of the thrust integrated spectrum from resummation variations in both the standard and $\sigma\text{-improved}$ schemes and from fixed order variations.}
\label{fig:cumulanterror}
\end{figure}

\section{Example: Thrust}
\label{sec:thrust}

In this section we apply the $\sigma\text{-improved}$ scheme to the resummed thrust distribution in $e^+e^-$ collisions~\cite{Farhi:1977sg}. We define thrust as
\begin{equation}\label{eq:thrust}
\tau=1-\underset{\vec{n}}{\max}\,\frac{\sum_i|\vec{p}_i\cdot\vec{n}|}{\sum_i|\vec{p}_i|},
\end{equation}
where $\vec{p}_i$ are the three-momenta of the particles in the event and the maximization over unit three-vectors $\vec{n}$ determines the thrust axis.
The limit $\tau\to 0$ corresponds to two collimated back-to-back jets. The distribution of $\tau$ depends on different energy scales such as the collision center of mass energy $Q$, the typical jet mass $Q\sqrt{\tau}$, and the typical energy of soft emissions $Q\tau$. As discussed in Sec.~\ref{sec:formalism}, logarithms of ratios of these scales appear in the fixed order prediction of the spectrum and cumulant, and, near the threshold region $\tau\to 0$, these logarithms become large and have to be resummed.
 
Figure~\ref{fig:cumulantcompareSTD} shows the integrated $\sigma\text{-improved}$ spectrum at NLL$'$+NLO and NNLL$'$+NNLO \cite{Fleming:2007qr,Schwartz:2007ib}. The $\sigma\text{-improved}$ spectrum is consistent with the cumulant: it integrates to the inclusive cross section and reproduces its uncertainty. Convergence is also preserved in the new scheme as shown in \Fig{fig:converge}. The upper panels show the peak and transition regions of the $\sigma\text{-improved}$ spectrum at NLL$'$+NLO and NNLL$'$+NNLO. The lower panels show a comparison between the NLL$'$+NLO and NNLL$'$+NNLO predictions of the cumulant and of the integrated $\sigma\text{-improved}$ spectrum. These are the same curves in \Fig{fig:cumulantcompareSTD} but are reproduced here to emphasize that the integrated $\sigma\text{-improved}$ spectrum has the convergence properties of the cumulant.

We stress that the point-by-point uncertainties in the spectrum are equivalent to the standard case since the profile variations obtained with the Bolzano algorithm fill the standard fiducial band. However, in contrast to the standard case, each profile variation obtained with the Bolzano algorithm preserves the inclusive cross section, and thus encodes correlations between uncertainties at different points of the spectrum. This leads to the significant difference in uncertainties between the standard and $\sigma\text{-improved}$ integrated spectra shown in \Fig{fig:cumulantcompareSTD}. We illustrate this further in Fig.~\ref{fig:cumulanterror} by comparing the uncertainty $\Delta(\tau)$ of the integrated spectrum from the resummation variations in both schemes as well as from the fixed order variations. 

As expected, the resummation uncertainty in the $\sigma\text{-improved}$ scheme goes to zero at large $\tau$, leaving only the fixed order uncertainty. Thus, in the $\sigma\text{-improved}$ scheme, the resummation uncertainty is fully uncorrelated with the inclusive cross section, while the fixed order uncertainty is fully correlated with the inclusive cross section. This makes it straightforward to build a covariance matrix from uncertainties in different bins of the observable. 
For example, in the case of two bins, the resummation and fixed order uncertainties exactly map into migration and yield uncertainties defined in \Ref{Stewart:2011cf} (see also Refs.~\cite{Stewart:2013faa,Gangal:2013nxa}), and they are given in Fig.~\ref{fig:cumulanterror}  for the 2-jet bin defined by the interval $[0, \tau]$.

\section{Conclusions}
\label{sec:conclusions}
Resummed predictions for cumulants and spectra of generic QCD observables are often inconsistent. Even though the difference is formally higher-order, it can be numerically relevant, and furthermore, uncertainty correlations across the spectrum are not properly included in standard resummation schemes.

In this paper, we defined the $\sigma\text{-improved}$ scheme, a two-step procedure that makes cumulants, spectra, and their uncertainties consistent. In the first step, we provided a prescription to add higher-order terms that make the value for the integrated spectrum consistent with the inclusive cross section. In the second step, we devised Bolzano's algorithm to select profile scales that preserve the inclusive cross section, thus encoding proper uncertainty correlations across the spectrum. We applied the scheme to the thrust distribution at NLL$'$+NLO and NNLL$'$+NNLO, demonstrating consistent predictions for the cumulant and the integrated spectrum (\Fig{fig:cumulantcompareSTD}), good convergence properties (\Fig{fig:converge}), and robust uncertainty estimation (\Fig{fig:cumulanterror}). As discussed in \Sec{sec:thrust}, in the $\sigma\text{-improved}$ scheme, resummation and fixed order uncertainties exactly map onto migration and yield uncertainties, and thus a covariance for different bins of an observable can be straightforwardly computed~\cite{Stewart:2011cf}.

The $\sigma\text{-improved}$ scheme defines a general strategy which can be applied to other observables. For example, it would be interesting to consider the Higgs transverse momentum and $C-$parameter distributions~\cite{Hoang:2014wka,Hoang:2015hka,Becher:2012yn,deFlorian:2011xf,Chiu:2012ir,Neill:2015roa}. Furthermore, we have implemented a simple version of Bolzano's algorithm; improvements to the computational efficiency and to the matching of the integrated spectrum and cumulant (e.g., by matching at additional intermediate points) can be pursued. In this paper, we worked with resummation within the SCET framework, but the same techniques can be applied to resummed spectra in dQCD.

\section*{Acknowledgments}
We thank Frank Tackmann for collaboration in the early stages of this work, as well as Christopher Lee and Iain Stewart for useful discussions. This work used resources of the National Energy Research Scientific Computing Center and was supported by the DOE under Contract No. DE-AC02-05CH11231. 


\appendix
\section{Implementation Details}
\label{app:Bolzano}

In this appendix we discuss an implementation of the $\sigma\text{-improved}$ scheme for the thrust distribution.

We first added higher-order terms to the thrust spectrum to restore consistency between its integral and the inclusive cross section, according to \Eqs{eq:higherorder}{eq:sigmaimproved}. The explicit form of the suppression factor $\kappa(\tau)$ we used is given by
\begin{align}
\kappa(\tau)_\text{NLL$'$}& =0.90625(1-\tanh(8\tau-0.56))\,,\\
\kappa(\tau)_\text{NNLL$'$} &=0.8475(1-\tanh(8\tau-0.56))\,.
\end{align}

For the profiles $\widetilde{\mu}_{S,J}(\tau)$ in the higher-order terms in \Eq{eq:higherorder}, we used the form
\begin{align}
\widetilde{\mu}_{S}(\tau)&=0.003+0.4985(1+\tanh(10\tau-2))\,,\\
\widetilde{\mu}_{J}(\tau)&=\sqrt{Q\widetilde{\mu}_{S}(\tau)}\,,
\end{align}
where $Q$ is the center of mass energy.

We now discuss Bolzano's algorithm, based on the solution presented in \Sec{subsec:Bolzano}. The goal is to fill a band, defined by boundary functions $\mu_i^{\rm min}(\tau)$ and $\mu_i^{\rm max}(\tau)$, with profiles whose integrated spectrum is equal (within tolerances) to the inclusive cross section. Additionally, we require that the profiles are monotonic, smooth (at least $C^1$), and have fixed shapes near the endpoints.

As mentioned in \Sec{subsec:Bolzano}, we will solve \Eq{eq:main} in the subdomain $[\tau_{\rm NP} , \infty)$, assuming that the contribution to the total cross section from $[0,\tau_\text{NP})$ is given by the cumulant. Thus, \Eq{eq:main} is replaced by
\begin{align}\label{eq:main2}
\sigma_{\rm incl} = \Sigma \big(\tau_\text{NP};\mu_i (\tau_\text{NP})\big) + \int_{\tau_\text{NP}}^{\infty} \df \tau \frac{\df\sigma_R}{\df\tau} \big( \mu_i\big).
\end{align}
The shape of profiles in the region $\tau\leq \tau_\text{NP}$ is fixed by the value at $\tau_\text{NP}$ as
\be
\label{eq:muNP}
\mu_i(\tau) = {\mu_i(\tau_\text{NP}) \over \mu_c(\tau_\text{NP})}\,\mu_c(\tau) \, ,\quad 0\leq\tau<\tau_\text{NP},
\ee
so that the cumulant depends on $\mu_i(\tau_\text{NP})$. Above, $\mu_c(\tau)$ is the default central profile, here taken to be the standard central profile. In general, \Eq{eq:main2} is no longer linear in the spectrum, and the solution in \Eq{eq:alpha0} is invalid. However, if we consider profiles with the same value at $\tau_\text{NP}$, the boundary term is simply a constant, and we recover, with appropriate redefinition, the linear system solved by \Eq{eq:alpha0}. The complete set of solutions that fill the bounding functions can then be obtained by considering various values of $\mu_i(\tau_\text{NP})$.

The first step is to generate profiles, and it is done by interpolating a curve through randomly sampled points. Gridded sampling can also be done, but may be less efficient given the unknown distribution of solutions and the high dimensionality required for coverage. To obtain solutions with the properties listed below \Eq{eq:main}, the generated profiles have the following properties:
\begin{enumerate}
\item Fixed shapes near endpoints. In particular, all profiles have the same value at $\tau_\text{NP}$.
\item Roughly bounded: $\mu_i^{\rm min} \lesssim \mu_i \lesssim \mu_i^{\rm max}$.
\item Roughly monotonic: $0 \lesssim {\df \mu_i \over \df \tau } $.
\item Smooth and artifact-free.
\end{enumerate}

The first property enforces standard behavior in the non-perturbative and tail regions, as discussed in \Sec{subsec:Bolzano}. We require that solutions obtained in the subdomain connect smoothly at $\tau_{\rm NP}$ to the functional form given in \Eq{eq:muNP}. In the tail region, imposing strict boundedness,
\be
\mu_i^{\rm min}(\tau) \leq \mu_i(\tau) \leq \mu_i^{\rm max}(\tau)\, ,\quad \tau>\tau_\text{tail},
\ee
is sufficient for obtaining solutions that reduce to unity since both boundary functions obey $\mu_i^{\rm min,max}(\tau) \to 1$. Note that in standard approaches, additional uncertainty is accounted for by varying the value of $\tau_{\rm tail}$. Here, this uncertainty is effectively accounted for by appropriate choice of the boundary function $\mu_i^{\rm max}(\tau)$.

The second property follows from the $\mu$-range of the chosen sampling regions. Note that profiles that are slightly unbounded or non-monotonic may still lead to solution profiles that are bounded and monotonic. For example, considering profiles outside the boundary functions may be useful for obtaining solutions close to (and within) the boundaries. On the other hand, profiles that are too far outside the bounds, or are highly oscillatory, are unlikely to yield acceptable solutions, and are not considered. 

The third and fourth properties require a careful choice of the $\tau$-range of sampling regions, and of the interpolation method employed. For example, we may avoid highly oscillatory profiles by controlling the number, range and locations of the sampling regions, or by using a monotone interpolation method. Similarly, we are careful to choose sampling and interpolation strategies that do not lead to profiles that exhibit artifacts such as kinks, nodes, and gaps.

Steps two through five described at the end of \Sec{subsec:Bolzano} are straightforward and we will not discuss them further.

In our basic implementation of the algorithm, it takes ${\cal O} (0.01)$ seconds on a standard laptop to generate one profile solution. We also obtain similar efficiencies when solving directly for the profiles using \Eq{eq:linear2} and a numerical root-finding method. Note that the efficiency depends not only on the solver but also on the strategy for generating initial candidate profiles, and the guidelines for step one described above are useful to increase the yield of solutions with the required properties. Interestingly, there may be small regions within the boundary functions that are hard to fill with solutions, demanding precise correlations at small and large $\tau$ to yield the right cross section while maintaining monotonicity. These gaps can be supplemented by focusing the initial sampling regions, but in practice, the remaining gaps are not large enough to significantly impact the point-by-point uncertainty estimation.

\bibliography{bolzano}

\end{document}